\documentclass[sigplan,nonacm]{acmart}
\usepackage{enumitem}
\usepackage[T1]{fontenc}
\usepackage{textcomp}
\usepackage{xcolor,booktabs,graphicx,amsmath,hyperref}
\newcommand{\syn}[1]{\textcolor{red}{#1}}
\newcommand{\meas}[1]{\textcolor{blue}{#1}}

\newcommand{\us}{\textmu s}
\title{Agora: A System for Consumption-Based GPU Pricing}

\author{Ian McDougall}
\email{imcdougall@wisc.edu}
\affiliation{%
  \institution{University of Wisconsin-Madison}
  \city{Madison}
  \state{WI}
  \country{USA}
}

\author{Noah Scott}
\email{nmscott@wisc.edu}
\affiliation{%
  \institution{University of Wisconsin-Madison}
  \city{Madison}
  \state{WI}
  \country{USA}
}

\author{Karthikeyan Sankaralingam}
\email{karu@cs.wisc.edu}
\affiliation{%
  \institution{NVIDIA \& University of Wisconsin-Madison}
  \city{Madison}
  \state{WI}
  \country{USA}
}

\begin{document}
\begin{abstract}
We present Agora, a lightweight system designed for cloud-based GPUs which meters a three-dimensional resource vector from commodity performance counters and prices it with a function that charges customers based on resources used rather than time spent renting. Because the calibration is revenue-neutral, any gain must come from demand the flat rate had priced out, and we report under what conditions this consumption-based pricing scheme increases revenue: we find that the switch is revenue-negative at the median on a fixed market, but provides benefits in 86\% of cases where a provider serves 15–30\% of its addressable market against 10\% where it serves over 60\%. Revenue expands dramatically in proportion to the number of customers priced out by the flat rate: up to 141\% in an optimistic case.
\end{abstract}
\maketitle
\section{Introduction}
\label{sec:intro}

GPU compute is sold in units of time: a tenant rents an
accelerator, or a node of them, by the second or the hour. Time is not what
a workload consumes. A modern accelerator is a bundle of separately
consumable resources---tensor throughput, memory bandwidth,
interconnect---and two tenants renting the same node-hour can stress
entirely different ones. For example, a latency-sensitive decode workload saturates HBM
bandwidth while leaving the tensor cores nearly idle; a long-context prefill
does the opposite, on the same model and the same hardware. Time-based pricing charges both the same, so the
resource-intensive tenant subsidizes the moderate one and neither price
reflects what either consumed. Splitwise~\cite{splitwise} already routes the
two phases to different hardware because one machine at one price serves
both inefficiently---a workaround for a pricing model that cannot see the
resource a workload consumes.

The disconnect is not marginal. On a single class of hardware, the workloads
a provider serves diverge in resource consumption by more than an order of
magnitude (\S\ref{sec:heterogeneity}), so a flat hourly rate
systematically over-charges light workloads and under-charges heavy ones
with no hardware difference to justify it. That heterogeneity is now an
economic force and not only a technical observation: in mid-2026 Nvidia
structured a \$500\,billion data-center financing package around a
guarantee that aging accelerators retain resale value, on the explicit
premise that when one tenant's needs change the same hardware can be handed
to ``another customer, another cloud, or another
operator''~\cite{nvidia-residual}---the cross-generational shadow of the
within-generation heterogeneity we measure.

Pricing has begun to move as well: cloud providers have broken the
flat-rate paradigm with per-second billing, per-token pricing and
provisioned-throughput tiers (\S\ref{sec:motivation-survey}).
\emph{Feature-based pricing} (FBP)---billing a tenant for the measured
resource vector its workload consumes rather than for the wall-clock time it
holds the hardware---is the principled endpoint of that transition, and, we
argue, a profitable move for the providers positioned to make it
(\S\ref{sec:incentive}). We use ``feature-based pricing'' and ``consumption-based''
interchangeably throughout.

The natural and somewhat traditional alternative is to price hardware
\emph{tiers}: offer fast and slow hardware at different rates and let
time-sensitive buyers pay for speed, which recent mechanism-design work
shows recovers much of the value a single flat rate leaves on the
table~\cite{ec26}. But tiering prices the \emph{hardware}, not the
\emph{consumption}. Even on one class of hardware at one hourly rate,
workloads differ by up to $45\times$ on individual resource
dimensions (\S\ref{sec:heterogeneity}), and a flat rate bills them
identically. Tiering
fixes which hardware a buyer lands on; consumption pricing addresses how
differently two buyers use that hardware once there (\S\ref{sec:market}).

We present Agora, a system for consumption-based GPU pricing. Agora meters
each rented GPU's resource-utilization vector from hardware counters, prices
it with a function calibrated to be revenue-neutral against the workloads being run on a given cloud provider's platform, and retains the per-sample record as the itemized
justification for the resulting charge---a bill a tenant can read against
its own workload rather than take on faith. We demonstrate that Agora is a
competitive instrument in the flexible, price-sensitive segment of the cloud market.

We first introduce the concept of feature-based pricing, and whom it benefits (\S\ref{sec:motivation}). Then we describe the resource-vector (\S\ref{sec:vector}) and pricing functions (\S\ref{sec:mechanism}) we use to realize FBP, followed by a description of \textit{Agora}, the system architecture used to implement it for real users (\S\ref{sec:system}). Finally, we evaluate FBP against a flat-rate across a corpus of LLM workloads (\S\ref{sec:eval}), and explore the use of FBP beyond the LLM use-case (\S\ref{sec:torchbench}).


This paper makes five contributions:
\begin{itemize}[leftmargin=*]
\item \textbf{An empirical characterization of GPU resource heterogeneity}
(\S\ref{sec:eval}). On a single 8$\times$B200 node, across eight dense and
mixture-of-experts models at a range of batch sizes and input/output context
lengths, time integrated resource consumption varies by up to
$45\times$ on individual dimensions; and the profile depends
jointly on model and operating point, so no static rate keyed on
model or hardware can price it.

\item \textbf{A resource-vector pricing mechanism} (\S\ref{sec:mechanism})
that is revenue-neutral by calibration, clipped into a predictability band
the tenant knows before renting, and non-decreasing in every resource---so a
tenant can never lower its bill by manufacturing consumption.

\item \textbf{Agora} (\S\ref{sec:system}), a metering, pricing and logging
pipeline. We bound ingestion cost at
400--800-GPU scale and find that fan-in rather than sampling
frequency is what binds (\S\ref{sec:ingestion}).

\item \textbf{An incentive analysis} (\S\ref{sec:incentive},
\S\ref{sec:market}) of why a provider would adopt consumption pricing. We measure this rather than model it: every configuration in the workload corpus is priced individually, a willingness-to-pay (WTP) distribution decides who rents under each regime, and we sweep that distribution, the job mix and the size of the addressable market. We find that across our sweep the provider increases their revenue in 66\% of configurations. This increase can be anywhere from 1.9\% to 416\%, depending on how many customers they are able to attract and how willing resource-intensive customers are to pay a premium.

\item \textbf{A stress test outside the design corpus}
(\S\ref{sec:torchbench}). We apply the calibrated mechanism, unchanged, to
single-GPU applications, which generate no interconnect traffic and load the
two remaining axes quite differently from the serving corpus, and report
where its calibration holds and where it breaks.

\end{itemize}

\section{Consumption-Based Pricing}
\label{sec:motivation}

Time-based pricing was a reasonable approximation when accelerators were
balanced and workloads homogeneous. Neither holds today. This section shows
that the market has already begun pricing more granularly
(\S\ref{sec:motivation-survey}) and describes who has an incentive to switch to consumption-based pricing (\S\ref{sec:incentive}). Appendix~\ref{app:notation} lists every symbol
this section uses.

\subsection{The Market Has Already Left Flat Pricing}
\label{sec:motivation-survey}

Consumption-based pricing continues a trajectory the cloud market is
visibly on: flat hourly rental has given way to per-second rental, provisioned-throughput tiers that charge for the
right to lower latency, and per-token pricing in inference settings. We surveyed ten providers---six neoclouds (specialized cloud computing providers) and four conventional hyperscalers(\cite{coreweave, lambda, runpod, together, fireworks, modal, aws, gcp, azure, oci})---along billing granularity,
commitment model, differentiation, and any explicit consumption element
(Appendix~\ref{app:survey}). Several already meter GPU, CPU and memory \`a
la carte at per-second granularity. The pattern is not a clean
neocloud/hyperscaler split: granular pricing concentrates in the
\emph{self-serve, flexible} offerings on both sides, which is exactly the
price-sensitive, low-switching-cost segment where \S\ref{sec:incentive}
predicts a competitive wedge. The support is directional and structural
rather than a claim about revenue---list-price granularity does not reveal
how many dollars flow through each mode.

\subsection{Who Adopts Consumption-Based Pricing}
\label{sec:incentive}

Agora is calibrated so that, applied to a provider's
\emph{current} customers, it collects exactly the revenue of the flat rate it
replaces, and each bill is clipped into a band around the flat cost of the
same rental, since a customer may prefer a worse price it can predict to a
better one it cannot (\S\ref{sec:price-family}). On day one, then, no average
bill changes---and clipping is not free, since truncating the largest bills
hands back revenue the neutral calibration had counted on. We write $\kappa$
for that give-back. A provider asked to instrument its fleet for a slightly
negative wash is entitled to ask what it is buying.

\subsubsection{Exploiting Uniform Pricing}
\label{sec:cream}

Resource-intensive customers cost more to serve: they draw more power and,
in any multiplexed setting---MIG, fractional-GPU, serverless pools---tie up
capacity that could be sold to another tenant. A flat rate charges both the
same for an hour, so per-customer margin \emph{decreases} in resource
intensity. That is an opening a competitor can exploit without refusing
anyone: offered a feature-based rate, light customers switch and heavy ones
stay, leaving the incumbent the expensive half of its own pool
(Appendix~\ref{app:cream}). Consumption pricing is not a concession to
customers; it is how a challenger sorts an incumbent's pool.

\subsubsection{When the wedge is revenue-positive}
\label{sec:breakeven}

Consumption pricing redistributes bills: light workloads pay less and heavy
ones more, and on a fixed base the two cancel up to $\kappa$. The base does
not stay fixed---heavy customers may leave, light ones may arrive---and the
direction of that race follows from the calibration alone, without any
assumption about buyers. Revenue on a fixed base is the flat rate's revenue
minus $\kappa$, so revenue can move only through \emph{quantity}, and the
premium collected from heavy customers is not new money: neutral calibration
has already committed it to the light segment's discount. Any gain must come
from arrivals. If none do, FBP is revenue-negative by exactly $\kappa$, and
by more if any heavy customer leaves---so the cap sets a bar expansion must
clear before the mechanism gains anything at all. \S\ref{sec:market} measures
where that bar is cleared, pricing every measured configuration and letting a
willingness-to-pay distribution decide who rents under each regime.

\paragraph{Two Caveats} A provider decides on profit, not revenue, and
serving joiners costs something. The sign of that gap is a capacity question
rather than a property of the mechanism, and at zero churn it is
unconditionally unfavourable (Appendix~\ref{app:profit}). Separately, the
segments here are binned by \emph{bill} while \S\ref{sec:cream} orders
customers by \emph{cost}, so skimming requires that cheaper-to-serve
customers in fact receive smaller bills, which is a condition on the rate vector we assume. 

\subsubsection{Why the incentive is asymmetric}
\label{sec:asymmetric}

The same reasoning says \emph{for whom} FBP pays, and the answer does not map
onto provider categories: the asymmetry runs between workload segments, which
cut through neoclouds and hyperscalers alike. Frontier labs and sustained
training buy negotiated capacity priced by the deal rather than the meter and
do not expand on a marginal price cut, so there are no joiners to gain.
Developer experimentation and bursty inference are price-sensitive and weakly
attached---and that is where granular pricing has in fact emerged, on both
sides of the divide (\S\ref{sec:motivation-survey}).

Three scope conditions follow: FBP incentives weaken on dedicated bare metal,
where the congestion channel vanishes and only power remains; where light
workloads are a small share of revenue; and where light demand is genuinely
inelastic. The last is binding and admits no qualifier---any gain depends on
attracting light expansion, and no amount of heavy inelasticity substitutes
for it.
\section{The Resource Vector}
\label{sec:vector}

Pricing consumption requires a vector that is simultaneously
\emph{economically meaningful} (the dimensions are separately scarce and
separately valued), \emph{statistically informative} (they carry
cross-workload variance, or pricing them is pointless), and
\emph{operationally measurable} (a billing system can sample them at line
rate from trustworthy counters). This section fixes that vector, states what
it omits, and describes the measurement path.

\paragraph{The resource vector.} We price three dimensions: \emph{tensor
throughput} (the rate of arithmetic the workload extracts from the tensor
units), \emph{HBM bandwidth} (the rate at which it moves data to and from
device memory), and \emph{NVLink traffic} (the rate at which it moves data
over the scale-up interconnect). These are the resources a workload can
independently bottleneck on: the roofline model already treats compute and
bandwidth as distinct binding constraints, and interconnect binds
independently when a model is sharded across GPUs. Each maps to a distinct
piece of the accelerator's bill of materials and to a distinct physical
scarcity, which is what makes pricing them separately meaningful rather than
arbitrary. Throughout, $D = 3$ is the number of priced dimensions and $u(t)
\in [0,1]^D$ the instantaneous vector of fraction-of-peak utilizations.

\paragraph{Measurement.} All three dimensions are exposed as hardware
counters by the data-center GPU telemetry interface: the
\texttt{tensor\_active} and \texttt{dram\_active} counters, and NVLink
byte counters normalized to link capacity~\cite{dcgm}. We read them through
DCGM at $\sim$10\,Hz ($\sim$100\,ms) to supply data for section \S\ref{sec:eval}; \S\ref{sec:torchbench} stress-tests the mechanism using similar metrics (excluding NVLink usage) collected via Nsight Compute.

A rate this coarse looks like a compromise but is not. The quantity
Eq.~\ref{eq:bill} prices is a time-integral, $\int u_r\,dt$, and DCGM
reports each duty cycle as a hardware time-average over its own sample
interval; summing duty cycle against interval therefore recovers that
integral up to counter error at any polling rate the interface supports. 
What it does cost, however, is \emph{resolution within} the rental rather than
accuracy of its total: a 100\,ms record cannot localize a charge to a phase
of the run, nor describe how sharply a collective loads the interconnect
while it is resident. That is an explainability limit
(\S\ref{sec:primitives}) rather than a pricing error, and would become one
only under a price sensitive to the shape of consumption rather than its
total. Separately, the counters carry a bounded
measurement error that both parties to a consumption-priced bill must accept
as the root of trust.

\section{The Pricing Mechanism}
\label{sec:mechanism}

This section converts measured consumption into bills under four
requirements: \emph{revenue targeting}, so a provider calibrates against its
current customers' resource usage; \emph{predictability},
since the customer-facing advantage of hourly pricing is a bill known in
advance; \emph{monotonicity}, so a customer can never lower its bill by
consuming \emph{more} of any resource, which is what forecloses
metric-faking; and \emph{measurability}, so every priced quantity is a
counter the system of \S\ref{sec:system} samples. The first two are
calibration properties (\S\ref{sec:price-family}), the third structural
(\S\ref{sec:robustness}), while the fourth was settled by \S\ref{sec:vector}. We do
not claim the stronger property that no tenant can lower its bill by
\emph{shifting} its resource mix; \S\ref{sec:robustness} explains why that
would be the wrong thing to want.

\subsection{The Price Function Family}
\label{sec:price-family}

Let $\bar p_g$ be the prevailing flat rate for a node-hour on tier $g$, and let $u(t) \in [0,1]^D$ be the sampled
utilization vector of \S\ref{sec:vector}. A bill has two parts: a fixed
charge for occupying the hardware at all, plus a usage charge summing each
resource's consumption times its per-unit price. Writing $\alpha_g$ for the
occupancy charge and $\beta_{g,r}$ for the per-unit price of resource $r$,
the bill for a window $[0,T]$ is
\begin{equation}
\mathrm{Bill} \;=\; \mathrm{clip}_{[\,\alpha,\,1.5\,]\cdot \bar p_g T}
\int_0^{T}\!\Bigl(\alpha_g + \sum_{r=1}^{D} \beta_{g,r}\, u_r(t)\Bigr)\,dt,
\label{eq:bill}
\end{equation}
accumulated per-sample by the node daemon (\S\ref{sec:system}) and clipped
once per window against the flat cost $\bar p_g T$. The integrand is the
instantaneous billing rate: $\alpha_g$ is what a tenant pays for holding the
hardware, and $\sum_r \beta_{g,r}u_r(t)$ what it pays for working it. Three
calibration levers follow.

\paragraph{Intercept.} $\alpha_g$ prices occupancy---host, chassis and idle
power are incurred regardless of utilization---and a strictly positive
intercept removes the squatting strategy in which a tenant holds reserved
hardware near-idle for nearly nothing. We set $\alpha_g = \alpha\,\bar p_g$
with $\alpha = 0.25$, and set the \emph{lower clip equal to $\alpha$}
rather than to an independent constant, which makes the floor non-binding: since every $\beta_{g,r}\ge0$ and $u_r\ge0$, a tenant's
pre-clip bill ratio is $\alpha + \sum_r \beta_{g,r}\bar u^{(i)}_r/\bar p_g
\ge \alpha$. A higher floor would bind on the lightest tenants, so the bill
redistribution would collect a net premium
from the light end rather than paying one out. Tying them costs a lever:
raising $\alpha$ blunts the consumption signal but narrows the exposure
band, so a provider cannot sharpen the price without widening the range of
bills a customer must budget for.

\paragraph{Rates.} Each dimension's rate is set by one calibration step,
\begin{equation}
\beta_{g,r} \;=\; \frac{(1-\alpha)\,\bar p_g}{D\,\bar u_{g,r}},
\label{eq:rates}
\end{equation}
where $\bar u_{g,r}$ is the reference-mix mean utilization of resource $r$; Appendix~\ref{app:rates} derives the formula and proves the neutrality
exact. That mean is the \emph{time-weighted} mean of $u_r(t)$,
\emph{pooled across every GPU in the rental}, over the provider's reference
mix (Appendix~\ref{app:stats}).

We weight the dimensions \emph{equally}: Eq.~\ref{eq:rates} asserts each priced resource is worth
exactly $(1-\alpha)/D$ of the node price at the reference mean, and there is
no weight-free price over a vector. Equal shares need no information beyond
the measured mix, where a cost-weighted alternative needs internal cost data
a customer might not be able to verify, and they are symmetric, so the weights cannot steer
demand toward whatever the provider has spare. 

One consequence of the normalization is intended and one is a hazard. A
dimension the reference mix uses lightly acquires a high per-unit rate, so a
tenant saturating a resource the rest of the mix barely touches pays sharply
for it---exactly the tenant the mix-mean price exists to identify. But the
rate \emph{diverges} as $\bar u_{g,r} \to 0$, and single-GPU workloads
generate no NVLink traffic at all, so a reference mix drawn from them would
bill any workload that utilizes NVLink without bound were it to join.

\paragraph{Cap and floor.} Bills are clipped to
$[\alpha, 1.5]\times\bar p_g T$, i.e.\ $[0.25,1.5]$ at our $\alpha$.
This is the predictability guarantee: exposure is bounded in a band known
before the rental begins, narrower in relative terms than spot variation
customers already tolerate. Since the floor never binds, the band's cost is
entirely the cap's.

That number is a property of the reference mix and not of the mechanism, and
the dependence is stronger than it looks. Because $\beta_{g,r}$ scales as
$1/\bar u_{g,r}$, a book of light workloads produces high per-unit rates,
which pushes heavier tenants into the cap and inflates the give-back. A
provider whose book is dominated by short, lightly-utilized rentals
therefore pays several percent of revenue for the predictability band
alone---enough to exceed any demand-response gain \S\ref{sec:market}
finds---and should either recalibrate on a representative mix or widen the
band.

\paragraph{Why Revenue-Neutral?} Setting $\alpha = 1$ and
every $\beta_{g,r} = 0$ recovers the flat rate exactly, so
Eq.~\ref{eq:bill} defines a family that contains incumbent pricing as a
special case. Revenue-neutral calibration is one point
in it and makes no claim to be the best one. We pin it there for two
reasons. It needs no demand data, only the provider's own measured mix,
where any optimized calibration needs additional internal heuristics or information available only to the provider. And it
is what separates the mechanism's effect from a price rise: a provider that
simply wants more revenue can raise $\bar p_g$, so unless the calibration is
pinned, a revenue gain cannot be attributed to pricing consumption rather
than to charging more.

\subsection{Robustness to Consumption Shifting}
\label{sec:robustness}

Can a tenant lower its bill by \emph{shifting} its resource mix rather than
by doing less work? The mechanism observes a resource vector, not the useful
work behind it, so it cannot certify that two configurations accomplish the
same work, but it doesn't need to. Two guarantees hold regardless of
intent, and one question stays open.

\paragraph{Monotonicity.} The bill is
non-decreasing in every priced dimension, since each $\beta_{g,r}\ge0$, so a
tenant can \emph{never} lower its bill by consuming \emph{more} of any
priced resource. That forecloses the most worrying class outright---padding
a resource with redundant work, inflating a counter to appear busy, any
strategy that lowers the bill by manufacturing consumption---with no appeal
to what the tenant is really doing, and it is the multi-resource
generalization of the anti-metric-faking requirement of the original
single-resource design. Separately, the clip bounds any tenant's bill to
$[\alpha,1.5]$ times the flat cost whatever mix it runs, so squatting cannot
approach free and the provider's per-rental exposure is bounded above.

\paragraph{Limitations.} We do \emph{not} claim tenants cannot
lower bills by changing resource mix, nor should we want to: a tenant that
genuinely runs on a lighter, cheaper-to-serve mix \emph{should} pay less.
The only behaviour worth calling gaming is a bill that falls with no
corresponding fall in the cost of serving the tenant, and whether that is
possible depends on how well the rates track the provider's marginal costs.
Ours are calibrated for revenue-neutrality, not cost-alignment, so we do not
claim they are gaming-proof in that stronger sense, and the same gap is
what makes the segment identification of \S\ref{sec:breakeven} conditional.

\paragraph{Cross-generational Pricing} Another consideration to be made regards cross-generational FBP: how should we update prices on existing hardware when a new product is released? One simple solution, which is analogous to the existing practice in time-based schemes, would be to simply apply a flat discount across all priced resources on the old hardware. We leave the implementation and exploration of such a scheme as future work.

\section{Agora System Design}
\label{sec:system}

\subsection{What Consumption Pricing Requires}
\label{sec:primitives}

\begin{figure}[tbp]
    \centering
    \includegraphics[width=1\linewidth]{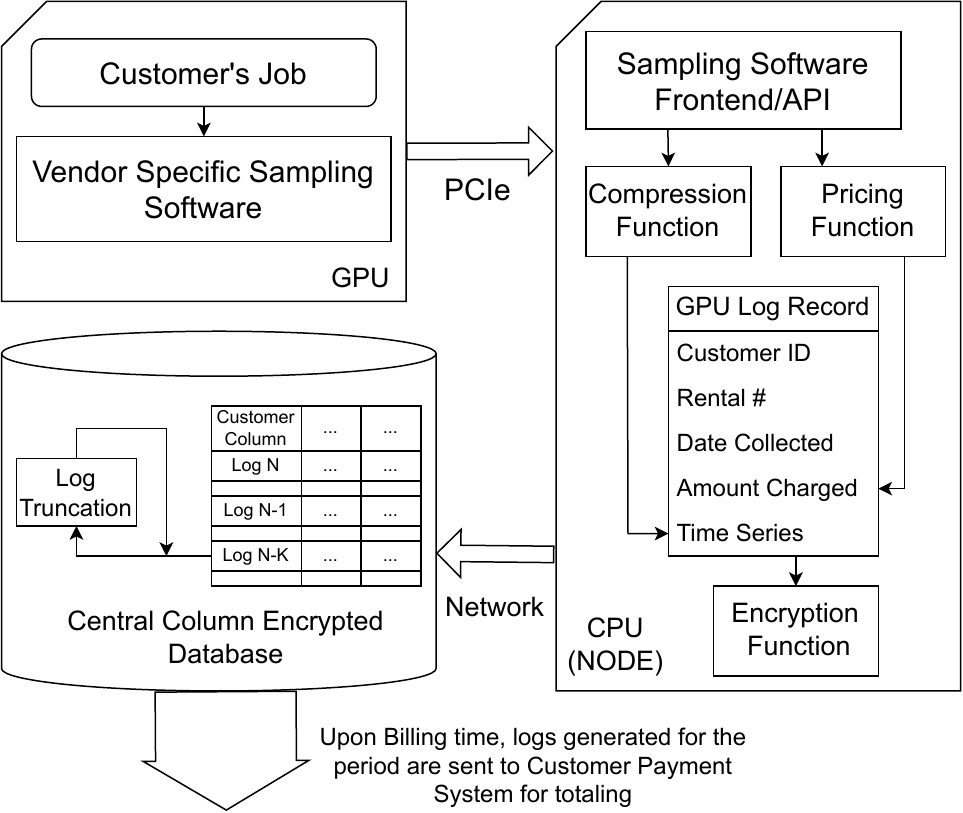}
    \caption{Proposed system overview.}
    \vspace{-0.2in}
    \label{fig:example_sys}
\end{figure}

Figure~\ref{fig:example_sys} sketches the architecture. Consumption-based
pricing asks two things of a provider that time-based pricing does not. It
must sample each rented GPU's resource counters at a regular rate for the
duration of every rental; and it must retain the resulting record in a form
that supports explainable, private billing. Time-based pricing needs
neither: its rate calculation observes no usage at all, only elapsed time,
which is why it can be operated with no telemetry infrastructure and why its
bills cannot be checked against anything.

The 8$\times$B200 systems of \S\ref{sec:method}, like the DGX-class systems
that preceded them, are sold in eight-GPU units; we refer to these and
similar systems as ``nodes.'' Each node collects performance counters from
each of its GPUs, places them in a log structure, and applies pricing,
compression, and encryption before the log is sent to a central server.

\paragraph{Sampling Rate Considerations.} Before
specifying a rate it is worth establishing what rate the price function
demands, because the answer is less than one might expect and it sets the
hardware Agora requires. Under a bill of the form
$\int_0^T(\alpha_g + \sum_r \beta_{g,r} u_r)\,dt$ (Eq.~\ref{eq:bill}),
replacing $u_r$ by its mean over each sampling window leaves
$\int_0^T u_r\,dt$ unchanged, and that integral is the only functional of
$u_r$ on which the bill depends. Period averaging therefore leaves the bill
\emph{exactly} unchanged---not approximately, and not on our corpus in
particular. 

The consequence is this section's central feasibility result. Because the
priced quantity is a time-integral, and because the telemetry interface
reports each duty cycle as a hardware time-average over its own sample
interval, summing duty cycle against interval recovers $\int_0^T u_r\,dt$ up
to counter error at \emph{any} polling rate the interface supports. Agora
therefore samples at the rate DCGM provides---$\sim$100\,ms in the
configuration available to us---and that is not a concession. An affine
consumption price is implementable on commodity telemetry today, with no
privileged counter access and no custom sampling stack: a substantially
weaker hardware requirement than a consumption-pricing proposal might be
assumed to carry.

\paragraph{Why a provider might still want finer resolution.} Sampling
faster does not change what a tenant owes, but it changes what the record
can \emph{show}. A bill computed from 100\,ms samples is exact in total and
mute about structure: it cannot tell a tenant which phase of its run drove
the charge, cannot localize a cost anomaly in time, and cannot support a
dispute at any granularity finer than the whole rental. Resolution is
therefore an explainability property rather than an accuracy requirement---
and it is also what a shape-sensitive price would require, were a provider
to adopt one. We size the pipeline against that
possibility rather than against present need.

\paragraph{Scalability.} The volume a
finer-grained deployment would generate is the binding design constraint, so
we bound it. Agora's sample record holds the three duty cycles of
\S\ref{sec:vector} rather than a single scalar; packed at 16-bit fixed point
per dimension with two bytes of metadata a sample still fits in 8 bytes, so
the three-dimensional vector costs no more to transport than a
single-resource design would. At the DCGM rate the resulting volume is
undemanding: 500 eight-GPU nodes emitting 8-byte records every
$\sim$100\,ms accrue roughly 10\,TB per year at
2.6\,Mb/s, which needs no special handling at all. Push the period to
50\,\us{}---the fastest rate we judge achievable on current hardware
at acceptable overhead, and the rate a shape-sensitive price would want---and
the same fleet would need more than 20\,petabytes per year (17.9\,PiB) at
5\,Gb/s. That 2000-fold span is why the architecture below is
built around truncation, compression, and hierarchical fan-in even though
the present operating point does not require them: we would rather specify a
pipeline that survives the envelope than one that only survives today's
telemetry. \S\ref{sec:ingestion} measures where in that span the pipeline
begins to bind.

\subsection{Architecture and Implementation}
\label{sec:architecture}

We propose a system, Agora, which is composed of three distinct parts: the GPUs and associated sampling software; customer-rentable nodes which manage collection, compression, pricing, logging, and encryption of sampled data metrics; and a central database server which collects the incoming logs. \textit{All of these components exist today}. We therefore present a prototype which illustrates how existing systems can be combined to support our novel pricing scheme, not a system whose architecture needs to be built from the ground up. 

After jobs are distributed to GPUs, usage is sampled and streamed to the
node's CPU for real-time pricing and compression. Calculated costs
accumulate in the log header, while usage data is appended to the time
series. Collected metrics are stored in a log as a compressed
time-data series, used primarily as a proof-of-charge item to be presented to
the customer at billing; this log also contains a header consisting of the
calculated amount to be charged, the date of logging, and the customer ID
and rental ID. 

Once a size limit is hit, logs are encrypted and transmitted to a
central database which collects the incoming logs and stores them in a ``rolling
frame'' database column, where the logs exceeding a generation threshold
$n$ are stripped of time-series data, leaving only headers. At billing
intervals, outstanding logs are forwarded to billing servers, with system
parameters adjusted based on bandwidth and storage availability. The
complexity of the compression function and the value of $n$ depend on the
facility's ability to transmit and store the volume of collected logs.

The retained time series is what makes a consumption bill explainable: a
customer receives not a duration and a rate but the metered vector its own
workload produced, itemized per dimension under Eq.~\ref{eq:bill}. We are
careful about what that does and does not establish. The record lets a
customer see \emph{how} a charge was computed and check it against its own
observations; it does not, on its own, prevent a provider from fabricating
the record. Making the log tamper-evident---by committing to it
incrementally in the style of a transparency log~\cite{merkle1987digital,laurie2013certificate}---is a
natural extension that the primitives here would support, and we treat it as
future work.

\paragraph{Sampling software.} At the operating point above, no custom
instrumentation is required: DCGM~\cite{dcgm} supplies the duty cycles
directly at acceptable overhead, which is the point of the feasibility
argument in \S\ref{sec:primitives}. A finer-grained deployment is a
different matter. NVIDIA's Nsight Systems~\cite{nsight} achieves sampling
periods below a tenth of a millisecond but carries tracing overhead that may
be unacceptable in a production multi-tenant setting, so a 50\,\us{}
deployment would require purpose-built sampling software. Such software
needs neither instrumentation nor tracing: it only needs to read the
performance counters, perform the necessary conversions, and store them into
a buffer so that the node can perform the rest of the logging procedures.
Whether any telemetry interface will expose counters at that rate at
tolerable overhead is an open question.

\section{Evaluating Feature-Based Pricing}
\label{sec:eval}

We evaluate FBP across a corpus of deep-learning workloads similar to those already served by granular pricing. Resource-based pricing is justified only if the workloads one provider
serves on one class of hardware, at one hourly rate, in fact consume the
underlying resources in sharply different proportions. We first confirm this fact in \S\ref{sec:heterogeneity}. Given this overall diversity, we then fix the choice of model and observe to what degree resource and bill diversity still exist (\S\ref{sec:dimensionality}). Otherwise, FBP could easily be reduced to a price based solely on model choice. We then perform a comprehensive sweep in \S\ref{sec:market} to determine how often and under what conditions FBP outperforms time-based pricing in terms of change in revenue. Finally, we investigate whether our proposed system architecture can handle ingesting pricing data from a large number of consumer nodes in a timely manner (\S\ref{sec:ingestion}).

\subsection{Methodology}
\label{sec:method}

\paragraph{Hardware.} All measurements are collected on a single
8$\times$B200 node (per GPU: 180\,GB HBM3e, 8\,TB/s memory bandwidth, 2250
BF16 TFLOPS, 1.8\,TB/s bidirectional NVLink). Confining the study to one
hardware generation isolates \emph{workload-driven} resource heterogeneity
from hardware-driven differences, so the mispricing we demonstrate cannot be
attributed to comparing dissimilar chips.

\paragraph{Deep-learning workload corpus.} We refer to each point within our corpus as a \textit{workload} or \textit{job}. Each workload consists of four values: the underlying LLM \textit{model}, its \textit{input context size}, \textit{output context size}, and \textit{batch size}. We include eight
models chosen to span the resource space granular pricing actually serves. Two
are dense: Llama-3-70B and Llama-3-405B. Six are
mixture-of-experts, at a range of active-parameter fractions: GPT-OSS-120B, GPT-OSS-20B, DeepSeek-V3.2
under expert parallelism, GLM-5, Qwen3-235B, and Qwen3-30B. The workloads are swept
over batch sizes $\{4,16,32\}$ and a grid of input and output context
lengths (input to 32\,K tokens, output to $\sim$64\,K,
excluding the jointly-long-context corner for tractability), with DeepSeek's
batch bounded by memory. The sweep is designed so that the compute/bandwidth
contrast can be observed either across models at fixed operating point or and across operating points within a model---rather than resting on a single comparison.

\paragraph{Workloads outside this corpus.} A mechanism validated only on
serving configurations would inherit their shape, so we reserve a second
class---single-GPU applications, which generate no interconnect traffic and
load the two remaining axes on a quite different scale---as a stress test of
the calibrated mechanism rather than folding it into the corpus the
mechanism is calibrated on. \S\ref{sec:torchbench} reports it separately for that reason, and
because it requires a different instrument.

\subsection{Cross-Workload Heterogeneity}
\label{sec:heterogeneity}

Every configuration in the corpus runs on the same 8$\times$B200 node and
would rent at the same hourly rate. Table~\ref{tab:dispersion} reports what
each of them consumes on the three priced axes, and what Eq.~\ref{eq:bill}
charges for it.

\begin{table}[t]
\centering\small
\begin{tabular}{@{}lrrrrr@{}}
\toprule
 & min & median & max & max/min & p95/p5 \\
\midrule
Tensor engine      & .0006 & .0210 & .1157 & 193 & 24.6 \\
HBM bandwidth      & .0036 & .1626 & .6429 & 181 & 45.5 \\
NVLink             & .0004 & .0092 & .0762 & 203 & 32.2 \\
\midrule
Bill / flat charge & 0.26 & 0.74 & 1.50 & 5.7 & 4.9 \\
\bottomrule
\end{tabular}
\caption{Time-integrated duty cycle per priced axis over 321
serving configurations on one 8$\times$B200 node, and the bill
Eq.~\ref{eq:bill} charges as a multiple of the flat charge for the same
elapsed time. The bill row is post-clip; its pre-clip maximum is
2.01. We use the \emph{uniform} job mix of
\S\ref{sec:market} for this table, which draws configurations with equal probability and
lets each contribute its own rented time.}
\vspace{-0.2in}
\label{tab:dispersion}
\end{table}

\paragraph{The axes are far apart.} Across 321 configurations,
workloads differ by $24.6\times$ on the tensor engine,
$45.5\times$ on HBM bandwidth and $32.2\times$ on NVLink,
measured as the ratio of the 95th to the 5th percentile. The extremes are
wider still, at $193\times$, $181\times$ and
$203\times$, but we quote the percentile figure throughout, because
the extremes measure something other than workload intensity. A single
30-second configuration holds the minimum on all three axes at once,
and it does so by being \emph{idle}: it works for about 2 of its
30 seconds, and over its active samples alone it is unremarkable,
reading $0.008$/$0.050$/$0.005$ against corpus medians
of $0.021$/$0.163$/$0.009$---below the median on every
axis, but nowhere near the $200\times$ its rental-averaged profile
implies. The corpus correlates resource activity with workload runtime---workloads with a duration below 60s are active, in the median case, for 29\% of the time; those above 600s are for 98\%---so a max/min ratio
here largely reports how much of a rental was used at all. Either measure settles what the mechanism needs
settled: on fixed hardware at a fixed rate, consumption is nowhere near
uniform.

\paragraph{Two axes co-move; the third does not.} Taken over the whole
corpus, all three pairs of axes correlate positively---Spearman correlation is 
$0.68$ for tensor against bandwidth, $0.51$ for tensor against
NVLink, and $0.67$ for bandwidth against NVLink. Read naively that
would undercut \S\ref{sec:vector}: axes that move together could be priced
by one of them. But note that a configuration that sits idle for most of its
rental reads low on all three, whatever it was running. Restricting to the
88 configurations longer than 600\,s, which are active for
$98\%$ of their rental and so carry no such common term, collapses
tensor against bandwidth to $0.01$ and tensor against NVLink to
$0.07$, while bandwidth against NVLink survives at $0.51$.
The apparent co-movement across the full
corpus is mostly shared idleness, not shared workload structure.

We therefore report the narrower claim. The memory and interconnect axes are
partly redundant on this corpus, and the tensor axis is independent of both,
so what justifies pricing a vector here is the axis that separates rather
than all three at once. \S\ref{sec:dimensionality} shows the same profile is
not recoverable from model identity either.

\textbf{Key Insight 1}: Across all configurations, workloads differ in their resource usage by as much as $45\times$. Once idle time is accounted for, we find that tensor and bandwidth decorrelate.

\subsection{Operating Point Moves the Profile}
\label{sec:dimensionality} 

If a workload's profile were fixed by its model identity, a provider could
price the model and be done. This is not the case, however. Holding the model set at Llama3-70B, we find that batch size is the dominant knob, and it moves the bill across the flat rate.
Going from batch 4 to 32 takes the median bill from
$0.81$ to $1.47$ times the flat charge for the same elapsed
time---an $82\%$ swing on one model, from comfortably below the flat
rate to well above it. The mechanism is the interconnect: NVLink duty rises
$4.1\times$ across that range as more tokens in flight enlarge every
collective, while HBM bandwidth rises only $8\%$ and the tensor axis
does not move monotonically at all. Input context barely moves the bill by
comparison, $0.88$ to $0.89$ from 256 to 32K
tokens, and it does so with tensor rising while HBM bandwidth falls and
NVLink does not move monotonically---so even the \emph{sign} of an axis
response depends on which knob is turned, and the axes that respond are not
the same ones.

Both knobs also lengthen the run, so both are exposed to the duty gradient
of \S\ref{sec:heterogeneity}, and we separate the two effects rather than
assume they are separate. Restricting to the 18 configurations
active for more than $95\%$ of their rental for this model, where that gradient
cannot operate, takes the batch swing from $82\%$ to $70\%$
and the context swing from $1.4\%$ to $0.6\%$. Roughly a sixth
of the batch effect is therefore attributable to idling and the rest to real work; the context effect is negligible under either
reading and we do not rest anything on it.

Two consequences matter for pricing. A provider charging by model identity
cannot separate these rentals, and neither can one charging by hardware
tier, since the entire spread sits on a single node type. And because the
effect crosses the flat rate rather than merely varying beneath it, the
\emph{direction} of the mispricing turns on the operating point: the same
tenant serving the same model is over-charged at low batch and
under-charged at high batch. Both are corrected under FBP.

\textbf{Key Insight 2}: Fixing the model does not stop the bill from moving. Increasing batch size takes the median bill from 0.81× to 1.47× the flat charge, which is directly correlated with a rise in NVLink usage.

\subsection{Comparing Feature- and Time-Based Pricing}
\label{sec:market}

The characterization above says the resource vector varies enough to be
worth pricing; it does not say that pricing it makes a provider better off.
\S\ref{sec:breakeven} narrowed that question to a single channel---any
revenue gain has to come from arrivals---without saying how large the
channel is. This subsection measures it.

\paragraph{Setup.} Each measured serving configuration---a model at a given
batch size and input/output context length---is one job, carrying its measured resource
vector, its duration, and hence both a flat charge and an FBP bill under
Eq.~\ref{eq:bill}. Configurations differ in duration, since output context
length sets how long a run takes, so revenue accumulates in dollars rather
than in job counts. Each job draws a willingness to pay as a multiple of the
flat rate, and under flat pricing it rents
iff that multiple is at least one. The incumbent set is therefore derived
rather than assumed: the $t{=}0$ state is an equilibrium of the flat regime,
and the two populations differ only in the way flat pricing itself
distinguishes them.

We calibrate the rate vector for revenue-neutrality over that incumbent set
alone and freeze it, then price the whole pool under FBP, where a job rents
iff its willingness to pay covers its bill. \emph{Leavers} are incumbents
whose premium pushes the bill above their WTP; \emph{joiners} are
non-incumbents whose discount brings it below a WTP the flat rate had
excluded. The reported quantity is the revenue of the post-switch served set
against that of the pre-switch one. Appendix~\ref{app:market} gives the
construction in full.

Joiners are absorbed rather than displacing incumbents, which assumes slack
at the margin. That is the regime in which attracting light demand is
possible at all, and it is consistent with the within-node contention
\S\ref{sec:cream} grounds its cost slope in. Where fleet capacity binds,
expansion becomes displacement and this construction does not apply; we note
only the direction, which favours the mechanism, since a provider choosing
which tenants receive scarce capacity is better served by a bill that tracks
consumed resources than by one that cannot distinguish them.

\paragraph{What is swept.} We sweep over two distinct modelling primitives: the customers' WTP and the nature of the addressable market. For WTP we vary the \textit{center} and \textit{spread} of the distribution, from 0.7 to 2.0 times the flat-rate, and 0.2 and 1.1, respectively. We also vary the distribution itself, choosing either lognormal, pareto, or uniform. Finally we vary the correlation between the WTP and intensity of the resource used, between 0.0 to 0.6. Positive correlation here means heavy jobs
absorb their premium because they value the compute, which dampens churn---so
independence is the conservative baseline and the correlated case is
reported beside it.

The second primitive is the size of the addressable market. Because
incumbency is derived rather than assumed, the joiner pool is by default
whatever the WTP distribution leaves priced out, and we call that the
\emph{self-consistent} case: it assumes nothing about latent demand. A
provider may of course face a market larger than its own book implies, so we
also sweep a multiplier on the price-excluded population from $1\times$ to $4\times$, and report the
\emph{book share}---how much of the addressable market the provider already
serves---as an axis in its own right. Both are assumptions about market size
rather than about pricing, so neither is folded into the headline. We also sweep
the job mix, which is not cosmetic: the mix sets the reference means, which
set the rates, which set how many bills reach the cap. Alongside the
categorical mixes---uniform, duration-weighted, short-job, long-job---we sweep a bell
curve over resource intensity, concentrating the fleet on lightly, moderately
or heavily utilizing configurations.

Each market is 10,000 job instances drawn with replacement from the
321 measured configurations, so a type recurs many times and each
instance carries its own willingness to pay. Increasing the job size to 100,000 agrees with the 10,000 instance sweep to within a point. It also does not matter if multiple jobs are being generated by the same customer - we merely report the cumulative effect of FBP across distinct workloads, not necessarily single-customer bills.

Revenue change is then reported against admitted new revenue at several
churn levels, so both axes are quantities the simulation counts rather than
parameters it infers. Two anchors bound the sweep. With no entry and no exit
the simulated change must come out at exactly $-\kappa$, which checks the
billing pipeline against the calibration and depends on no demand assumption
(Appendix~\ref{app:kappa}); with entry suppressed but churn allowed it is
strictly worse, which is the pure-downside case a provider should read
first.  

\begin{figure}[t]
\centering
\includegraphics[width=\columnwidth]{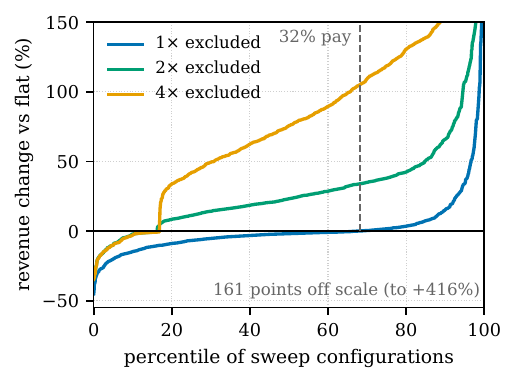}
\caption{Left: every configuration in the self-consistent sweep, sorted by
revenue change, one stem each; orange loses revenue, blue gains it, and the
dashed line is the crossing. Right: the same distribution as a percentile
curve at each assumed market size.}
\label{fig:marketsweep}
\end{figure}

\paragraph{Where the switch pays.} Across 3494 sweep points, drawn
in full in Figure~\ref{fig:marketsweep} and summarised in
Table~\ref{tab:market}, revenue change under FBP runs from $-45\%$ at
worst through $-15\%$ at the fifth percentile, $+18\%$ at the
median and $+146\%$ at the ninety-fifth, to $+416\%$ at best,
with $66\%$ of points revenue-positive. Nearly all of that upside is
one channel: how much of its addressable market the provider is \emph{not}
already serving. A provider holding 15--30\% of its market is
positive in 86\% of cases against 10\% for one holding over
60\%, and assuming price-excluded demand at $2\times$ or
$4\times$ what the corpus implies moves the median to $+23\%$
and $+76\%$ on $84\%$ and $83\%$ of points.
The only channel that is not a market-size assumption is the
correlation between willingness to pay and resource intensity, which works
through churn rather than entry: going from independence to
$\rho = 0.6$ lifts the median from $-4.2\%$ to $-0.2\%$
and the positive share from 20\% to 47\%, because heavy
tenants absorb their premium instead of leaving.

\begin{table}[t]
\centering\small
\begin{tabular}{@{}lrrrr@{}}
\toprule
 & \multicolumn{3}{c}{revenue change (\%)} & \\
\cmidrule(lr){2-4}
 & median & p5 & p95 & positive \\
\midrule
\multicolumn{5}{@{}l}{\emph{self-consistent market} (1197 points)}\\
\quad all points          & $-1.8$ & $-20$ & $+34$ & 32\% \\[2pt]
\multicolumn{5}{@{}l}{\quad\emph{by share of the market already served}}\\
\quad\quad over 60\%   & $-3.7$ & $-24$ & $+1$ & 10\% \\
\quad\quad 30--60\%    & $-0.7$ & $-15$ & $+13$ & 44\% \\
\quad\quad 15--30\%    & $+16.3$ & $-4$ & $+58$ & 86\% \\[2pt]
\multicolumn{5}{@{}l}{\quad\emph{by WTP--intensity correlation}}\\
\quad\quad independent        & $-4.2$ & $-22$ & $+25$ & 20\% \\
\quad\quad $\rho = 0.3$       & $-1.9$ & $-19$ & $+31$ & 28\% \\
\quad\quad $\rho = 0.6$       & $-0.2$ & $-16$ & $+51$ & 47\% \\
\midrule
\multicolumn{5}{@{}l}{\emph{sensitivity: larger addressable market}}\\
\quad $2\times$ excluded demand & $+23.1$ & $-10$ & $+107$ & 84\% \\
\quad $4\times$                 & $+75.7$ & $-10$ & $+229$ & 83\% \\
\bottomrule
\end{tabular}
\caption{Revenue change under FBP against flat pricing, over a sweep of
willingness-to-pay center, spread, family and intensity-correlation, job
mix, and addressable-market size. The self-consistent block assumes nothing
about latent demand; the sensitivity block does, and is never averaged into
the headline.}
\vspace{-0.25in}
\label{tab:market}
\end{table}

\paragraph{What it costs.} That headline averages
over assumptions about latent demand, so the self-consistent block---joiner
pool set to whatever the willingness-to-pay distribution leaves priced out,
assuming nothing---is the one to act on. Over its 1197 points revenue
change runs $-45\%$, $-20\%$, $-1.8\%$, $+34\%$
and $+289\%$ at the same five quantiles, and only $32\%$ of
points pay; the crossing in Figure~\ref{fig:marketsweep} sits at the
68th percentile, against the 16th and 17th once a larger
market is assumed. The accounting identity of \S\ref{sec:breakeven},
$\Delta R/R = e - \kappa - c\,L$, says why, with $e$ the revenue admitted
from joiners, $c$ the revenue that walked, both as shares of the incumbent
book, and $L$ the leavers' mean bill ratio. That $L$ lies in $(1,1.5]$ is
forced rather than measured---a leaver is an incumbent whose bill rose above
a willingness to pay that already covered the flat rate, and the cap bounds
every bill from above---so churn costs strictly more than the hours it
removes. Measured, $L$ runs $1.28$--$1.44$ at a median of
$1.34$, putting break-even at $5.6\%$ admitted entry against
$3.5\%$ churn and $18.7\%$ against $13.1\%$, with
$\kappa$ adding a median $1.1\%$ and up to $8.3\%$ on a
short-job book. At the 190 points where nobody joins and nobody
leaves both demand terms vanish and $\Delta R/R$ recovers $-\kappa$ to
machine precision, which checks the billing pipeline against the calibration
on no demand assumption at all (Appendix~\ref{app:kappa}). What this
licenses is therefore not that consumption pricing raises revenue: on a
fixed addressable market it is revenue-negative at the median, and the cases
where it pays are those where a provider has substantial unserved demand to
attract or its heaviest tenants are its least willing to leave.

\textbf{Key Insight 3}: FBP allows customers below the WTP range dictated by the flat rate to join. 86\% where the provider holds 15–30\% of its market against 10\% above 60\%. The change in revenue itself expands dramatically as the addressable market expands. 

\subsection{Does the Metering Pipeline Carry the Data?}
\label{sec:ingestion}

The pricing results above assume the metered record reaches the billing
server. At the DCGM operating point that is uninteresting---
\S\ref{sec:primitives} puts the fleet volume at 2.6\,Mb/s---but the
envelope a shape-sensitive price would demand is 2000-fold larger, and it is
the envelope that determines whether the architecture of
\S\ref{sec:architecture} is the right one.

\paragraph{Testbed.} Renting hundreds of GPU nodes to measure network
behaviour is impractical, so we emulate each node with eight
trace-replaying threads feeding the real logging, encryption (AES-256) and
transport path, hub-and-spoke to one central consumer that records each
arriving packet and its arrival time. Nodes and consumer run on separate
machines on a shared network, synchronized within a tenth of a millisecond
by Chrony with hardware timestamping. Runs cover 10k traces per GPU thread
at node counts of 50 and 100---400 and 800
emulated GPUs---and emission periods from 50\,\us{} to
1000\,\us{}, every one faster than the DCGM operating point by two
to three orders of magnitude. That is deliberate: the question is where the
pipeline breaks, not whether it survives today's load.

\begin{figure}[tbp]
    \centering
    \includegraphics[width=1.0\linewidth]{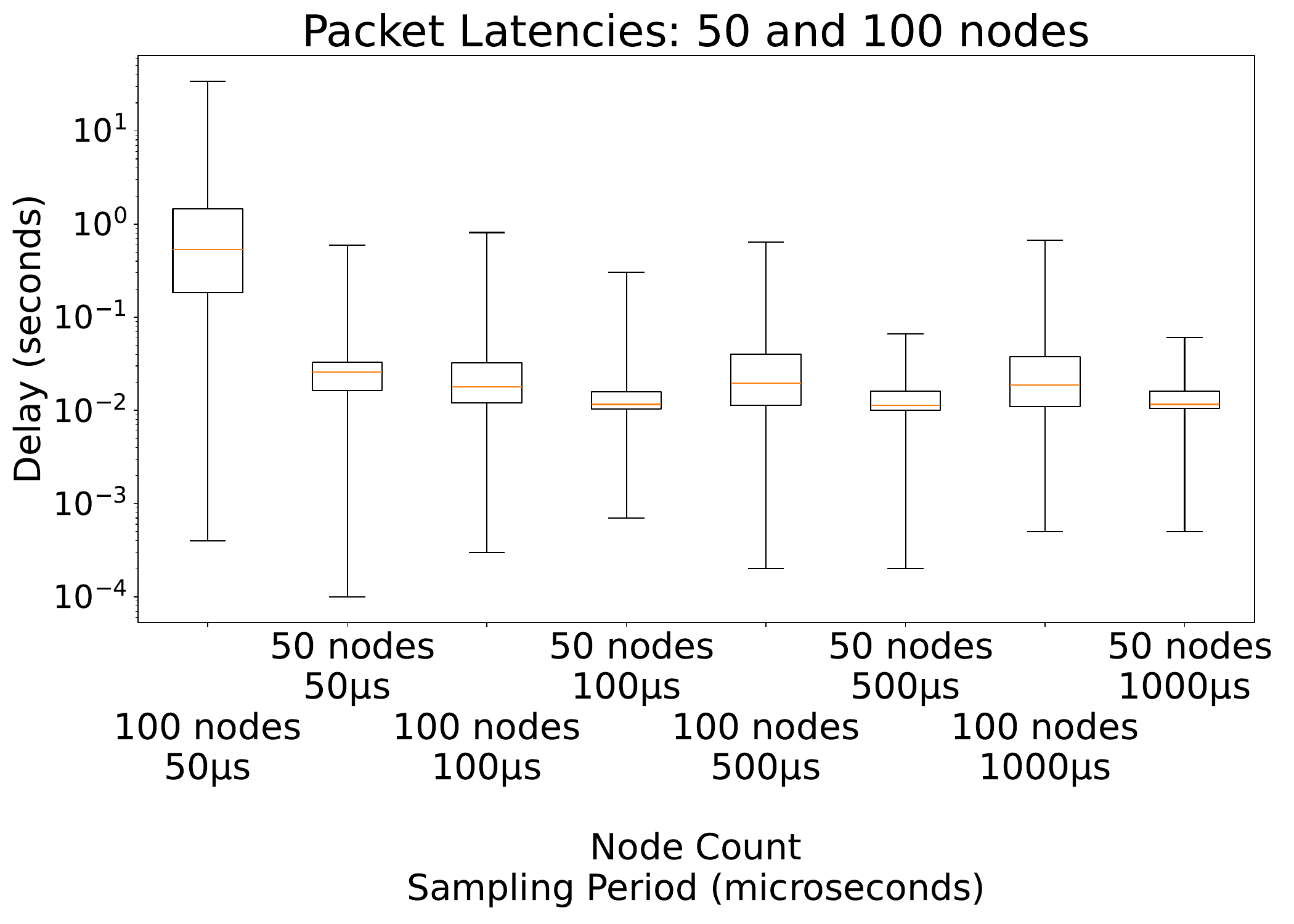}
    \caption{Producer-to-consumer log-transmission latency for 50 and 100
    producer nodes (400 and 800 emulated GPUs) over select emission periods.
    Emission rate barely moves the median and shifts only the tail;
    \emph{node count} is the binding pressure}
    \vspace{-0.2in}
    \label{fig:lat_boxplot}
\end{figure}

\paragraph{Ingestion is bounded by fan-in, not by emission rate.}
Figure~\ref{fig:lat_boxplot} reports two findings. Emission rate has little
effect on mean ingestion latency and moves only the worst-case tail, so
fine-grained sampling carries no proportional ingestion penalty. But
\emph{node count} is the binding pressure: at fixed emission rate, going
from 50 to 100 nodes worsens both mean and tail, most sharply
at the highest rate. The scaling constraint is fan-in rather than sampling
frequency, and it worsens with fleet size rather than pricing
granularity---so a deployment cannot read the first finding as permission to
sample fast at any scale. The mitigation, consistent with
\S\ref{sec:architecture}, is a hierarchical topology whose aggregators
consolidate a subtree's frames and forward during network lulls, trading a
tier of latency for fan-in relief.

The testbed replays telemetry rather than driving live
accelerators, so it measures the transport path and not GPU-side sampling
overhead, which we cannot measure without a counter interface that reads at
the 50\,\us{} envelope. And because every emission period swept sits
far above the DCGM operating point, this bounds the envelope rather than
characterizing the deployment: at $\sim$100\,ms the central consumer
is under no meaningful pressure at either node count.

\textbf{Key Insight 4}: Metering granularity is not what limits the pipeline. The binding constraint is fan-in, and it scales with fleet size rather than with pricing granularity. At the rate Agora actually samples, the consumer is never pressured at either fleet size.

\section{Pricing Beyond LLM Inference}
\label{sec:torchbench}

Everything to this point has been calibrated and evaluated on one workload
class: transformer inference served across a multi-GPU node. This
section applies the same mechanism, unchanged, to a class that violates most
of that corpus's regularities at once.

\paragraph{Methodology.} The corpus this section uses is 82 applications from the
TorchBench suite~\cite{torchbench}, collected on a single
H100---not the 8$\times$B200 node of \S\ref{sec:method}. Being
single-GPU, these applications sit at exactly zero on the interconnect
dimension, so we price $D=2$: the tensor engine and bandwidth. We measure
them per kernel with Nsight Compute (NCU)~\cite{nsight} rather than with DCGM,
for reasons of duration rather than pricing---the median application
accumulates 5.2\,ms of kernel time across 357 kernels, and
only 3 of 82 accumulate even 100\,ms, so a single
window at DCGM's $\sim$100\,ms interval would average an
application's entire kernel-active life together with a span of host-side
gap. This does not change the bill: the price of \S\ref{sec:mechanism} is
affine and recovers the same charge from a per-kernel trace as from any
coarser sampling of the same execution (\S\ref{sec:primitives}), subject to
one condition we enforce throughout---the trace is expanded onto the
wall-clock timeline of the rental, with host-side intervals between kernels
counted at $u=0$, since Eq.~\ref{eq:bill} integrates over the whole rental
and averaging over kernel-active time alone would overstate every profile
here. We make no claim comparing
this class's resource \emph{profiles} to the serving corpus's: the two
differ in workload, topology, hardware generation, instrument and time base
at once, and the instrumental differences are individually as large as any
workload effect. Every number below is internal to this class.

\begin{figure}[t]
\centering
\includegraphics[width=\columnwidth]{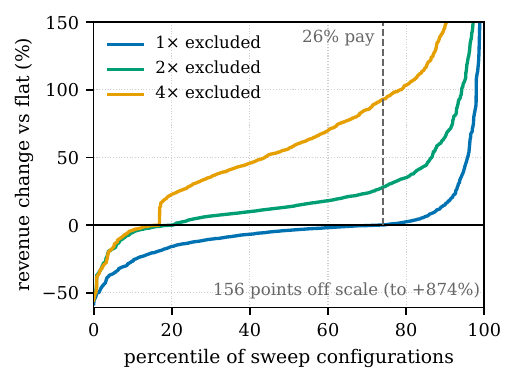}
\caption{The sweep of Table~\ref{tab:s7market}, drawn as
Figure~\ref{fig:marketsweep}.}
\vspace{-0.2in}
\label{fig:s7sweep}
\end{figure}

\paragraph{Where the switch pays.} We run the market experiment of
\S\ref{sec:market} unchanged---applications drawn with replacement into a
market of 10,000 instances, flat pricing deciding incumbency, rates
calibrated on the incumbent book alone and then frozen, everyone re-deciding
under Eq.~\ref{eq:bill}---over the same sweep of willingness-to-pay family,
centre, spread and intensity-correlation and addressable-market size, and an
analogous set of job mixes. Instead of long- or short-job only mixes used previously, we include \textit{training} and \textit{evaluation} only mixes. All other job mixes remain identical. Across 3494 points
(Table~\ref{tab:s7market}, drawn in full in Figure~\ref{fig:s7sweep})
revenue change runs from $-58\%$ at worst through $-28\%$ at
the fifth percentile, $+11\%$ at the median and $+141\%$ at
the ninety-fifth, to $+874\%$ at best, with $62\%$ of points
revenue-positive. The upside runs through one channel, how much of its
addressable market the provider is \emph{not} already serving: a provider
holding 15--30\% of its market is positive in 78\% of cases
against 6\% for one holding over 60\%, and price-excluded
demand at $2\times$ or $4\times$ what the corpus implies moves
the median to $+13.6\%$ and $+57.0\%$ on $80\%$ and
$83\%$ of points. Willingness to pay correlated with resource
intensity again works through churn rather than entry, lifting the median
from $-8.9\%$ to $-0.5\%$ and the positive share from
12\% to 43\%. On the self-consistent market, which assumes
nothing about latent demand, the same five quantiles are $-58\%$,
$-36\%$, $-3.8\%$, $+43\%$ and $+512\%$, and
$26\%$ of points pay; the identity of \S\ref{sec:breakeven} recovers
a leaver bill ratio of $1.32$ at the median, so entry has to beat
churn by about a third on this class too.

\begin{table}[t]
\centering\small
\begin{tabular}{@{}lrrrr@{}}
\toprule
 & \multicolumn{3}{c}{revenue change (\%)} & \\
\cmidrule(lr){2-4}
 & median & p5 & p95 & positive \\
\midrule
\multicolumn{5}{@{}l}{\emph{self-consistent market} (1199 points)}\\
\quad all points            & $-3.8$ & $-36$ & $+43$ & 26\% \\[2pt]
\multicolumn{5}{@{}l}{\quad\emph{by share of the market already served}}\\
\quad\quad over 60\%     & $-6.8$ & $-43$ & $+0.4$ & 6\% \\
\quad\quad 30--60\%      & $-3.3$ & $-25$ & $+15$ & 33\% \\
\quad\quad 15--30\%      & $+13.4$ & $-16$ & $+72$ & 78\% \\[2pt]
\multicolumn{5}{@{}l}{\quad\emph{by WTP--intensity correlation}}\\
\quad\quad independent          & $-8.9$ & $-38$ & $+12$ & 12\% \\
\quad\quad $\rho = 0.6$         & $-0.5$ & $-32$ & $+83$ & 43\% \\
\midrule
\multicolumn{5}{@{}l}{\emph{sensitivity: larger addressable market}}\\
\quad $2\times$ excluded demand & $+13.6$ & $-17$ & $+107$ & 80\% \\
\quad $4\times$                 & $+57.0$ & $-17$ & $+224$ & 83\% \\
\bottomrule
\end{tabular}
\caption{Revenue change under FBP against flat pricing on the single-GPU
class, over the same sweep as
\S\ref{sec:market}, over an analogous set of job mixes---the natural splits
of a single-GPU fleet, plus the same three intensity bells.}
\vspace{-0.25in}
\label{tab:s7market}
\end{table}

\paragraph{Calibration risks.} Calibration transfers
exactly. Reference means over this corpus are $0.073$ on the tensor
axis and $0.268$ on bandwidth, kernel-time weighted as duration
weights the serving mix, and the pre-clip mean bill over that book is
$1.000000$---Appendix~\ref{app:rates} holding on a class it was not
derived for, with the floor contributing nothing as
Appendix~\ref{app:kappa} requires. What distinguishes this class is not the
demand response but $\kappa$. Because $\beta_{g,r}\propto 1/\bar u_{g,r}$, a
book of lightly-utilising applications sets low reference means, raises
every rate and drives the heavy tail into the band: across the books this
sweep generates, $\kappa$ runs from $0.03\%$ of revenue on a
training-only book to $16.7\%$ on one concentrated in the lightest
applications, reaching $20.4\%$ at worst. At that extreme the
give-back exceeds the median admitted entry over the whole
sweep---$16.7\%$ against $8.9\%$---so the median light-book
provider loses about $\kappa$ by switching. It is not a universal loss:
45 of 169 light-book points still pay, but on entry shares
near $94\%$, which is a market-size assumption rather than a property
of the price. The band is also meaningful only over an integrated rental,
since priced individually the kernels inside a single application span
$11.8\times$ at the median and $21.6\times$ at worst before
clipping, with $24\%$ of them landing on a rail. A provider serving
this class should therefore recalibrate on a representative book or widen
the band (\S\ref{sec:price-family}): the predictability guarantee, not the
rate vector, is where its commercial risk sit.

\textbf{Key Insight 5}: FBP is robust when implemented against GPU workloads other than LLM inference. However, when the workload mix skews light, there is the risk of capping greater numbers of workloads under a naive calibration.

\section{Related Work}
\label{sec:related}

\paragraph{RaaS Cloud}
Similar resource-based cloud pricing schemes include \cite{agmon2014rise, ben2012resource}. This work proposes an alternative cloud model called the `Resource-as-a-Service' cloud (RaaS), in which computing resources (such as CPU cycles, memory frames, etc.) are bought and sold by cloud customers in an auction-style environment to optimize rental prices based on performance needs. To this end, an operating system \cite{ben2016nom} and various service frameworks \cite{agmon2014ginseng}, \cite{li2015rest}, \cite{luo2017adarm} have been proposed to be used within the RaaS cloud to automate and optimize fine-grained resource auctioning. Under RaaS, clients pay for hardware capacity, meaning they are billed even for idle resources. Our approach only bills resources actually used. 

\paragraph{Optimal pricing}
There is extensive prior work on optimal pricing in the operations research and microeconomics literature~\cite{myerson1981optimal,kleinberg2003value,besbes2009dynamic,den2014simultaneously,besbes2015surprising,cheung2017dynamic,misra2019dynamic,den2015dynamic}. 
Most of these works assume full knowledge of the buyer distribution, though several have also studied models that can approximate optimal prices under limited information~\cite{kleinberg2003value,besbes2009dynamic,den2014simultaneously,guo2023leveraging,kandasamy2020vcg}. 
While much of this work has been theoretical, recent applications have shown promising practical results~\cite{srivatsa2024deep,d2024disrupting,wang2024gemnet,ec26}.

\paragraph{Existing profiling and sampling solutions} Current research into frameworks and methodologies for data-center profiling include Propellor~\cite{shen2023propeller}, Accelerometer~\cite{sriraman2020accelerometer}, and Dmon~\cite{khan2021dmon}; these are not GPU focused nor do they examine the low-level hardware counters that are most useful to fine grain resource monitoring. Commercial and Industry tools generally assume the profiling entity has full control of a single-node or an entire cluster/data-center.  

Verifiable pricing is explored in the ALIBI system~\cite{10.1145/2517326.2451546} for CPU-centric cloud computing contexts. Existing work on high efficiency compression and long term storage of time-series data include Facebook's Gorilla~\cite{gorilla}, and UC Berkeley's BTrDB~\cite{btrdb}; Query-able Column encrypted databases have been explored in CryptDB~\cite{cryptdb} 

\section{Conclusion}
\label{sec:conclusion}

GPU compute is sold by time but consumed as a vector of separately
consumable resources. This paper characterizes that gap---on fixed hardware
at one hourly rate, workloads differ by up to $45\times$ on individual
dimensions, and the \emph{direction} of the mispricing turns on the operating
point---and closes it with a price that is affine in the metered vector,
revenue-neutral against the provider's existing customer-base, monotone in every
resource, and clipped to a band the customer knows before renting.
\emph{Agora} is the system that makes such a price deployable, collecting the
vector from commodity counters at line rate at an ingestion cost bounded by
fan-in rather than by sampling frequency.

Whether a provider should adopt it is conditional. Under revenue-neutral calibration any gain must come from demand the flat rate had priced out: on a
fixed market the switch is revenue-negative at the median, and it
pays where substantial demand is priced out, or where the tenants it charges
more are the least willing to leave. What time-based pricing made invisible,
consumption-based pricing makes legible.

\bibliographystyle{ACM-Reference-Format}
\bibliography{references,references2}

\appendix

\section{The Clipping Give-Back}
\label{app:kappa}

Revenue-neutral calibration (\S\ref{sec:price-family}) sets the rate vector
so that the mean pre-clip bill over the reference mix equals the flat rate.
Clipping perturbs that. Writing $\tau$ for the revenue the cap truncates and
$\phi$ for the revenue the floor adds back, the net give-back is
$\kappa \equiv \tau - \phi$: the amount by which a clipped mechanism falls
short of neutrality on a fixed customer base.

Two properties carry the argument of \S\ref{sec:breakeven}. First,
$\phi = 0$. Because the lower clip is set equal to the occupancy intercept
$\alpha$, a tenant's pre-clip bill ratio is
$\alpha + \sum_r \beta_{g,r}\bar u^{(i)}_r/\bar p_g \ge \alpha$ for its own
time-averaged utilization $\bar u^{(i)}_r$, since every $\beta_{g,r} \ge 0$
and every $u_r \ge 0$. The floor therefore never binds, and
$\kappa = \tau \ge 0$: the band can only cost revenue, never add it.
Second, $\kappa$ is the \emph{entire} cost of the predictability band on a
fixed base. An unclipped calibration is exactly neutral, so any fixed-base
shortfall is attributable to the cap and to nothing else, which is what
licenses \S\ref{sec:breakeven} to treat $-\kappa$ as the whole of the
no-response outcome.

Both properties are statements about the mechanism rather than about demand,
and \S\ref{sec:market} exploits that: with no entry and no exit the
simulated revenue change must come out at exactly $-\kappa$. That is a check
on the billing pipeline---rates, integration, clipping---against the
calibration, and it holds whatever the willingness-to-pay assumption.

Note finally that $\kappa$ is a property of the mechanism against a
\emph{mix}, not a constant. The reference mix sets $\bar u_{g,r}$, which
sets $\beta_{g,r}$, which sets how many raw bills exceed the cap, so
$\kappa$ must be recomputed whenever the corpus or the reference mix
changes.

\subsection{Revenue is not profit}
\label{app:profit}

\S\ref{sec:market} reports a revenue change; a provider decides on profit.
The two differ because demand response moves quantities and quantities cost
money to serve: departing heavy tenants take their serving cost with them,
and arriving light tenants bring theirs. We do not fold the difference into
the reported result, because it turns on segment margins this paper cannot
measure, but its sign is not arbitrary and three regimes are worth
separating.

At zero churn the correction is always unfavorable, for any
margins whatever: nobody leaves to save cost, while every joiner must still
be served, so the profit change is below the revenue change. Away from that
anchor the sign turns on a capacity question rather than on the pricing
mechanism. If the fleet must expand to serve joiners, those joiners bear
full average cost including depreciation and the revenue picture is
optimistic. If they instead fill capacity that would otherwise sit idle,
their short-run marginal cost is power rather than power plus depreciation,
the correction shrinks, and above modest churn it reverses---the provider
gains more profit than revenue, because the tenants it shed were the
expensive ones to serve.

Which regime a provider is in is therefore the first question to ask, and it
is not one this corpus can answer. \S\ref{sec:market} assumes slack at the
margin, which is the favourable case for this correction and also the only
regime in which attracting light demand is possible at all.

\section{The Cream-Skimming Schedule}
\label{app:cream}

This appendix supports the claim in \S\ref{sec:cream} that a revenue-neutral
consumption schedule sorts a uniformly-priced pool without having to refuse
anyone. Order customers by resource intensity $\theta$ and write the
provider's cost to serve an hour as $c(\theta) = c_0 + c_1\theta$: a fixed
part $c_0$ for powering and allocating the GPU at all, plus an
intensity-dependent part $c_1\theta$. The slope $c_1 > 0$ is the assumption
that heavier use genuinely costs more; its sources are power draw and,
dominant in any multiplexed setting, the opportunity cost of capacity denied
to co-tenants. On fully dedicated bare-metal $c_1$ is small, which is scope
condition~(i) in \S\ref{sec:asymmetric}.

\subsection{Self-selection versus profitability}
\label{app:selfselect}

Facing an increasing challenger schedule $p(\theta) = a + b\theta$, a
customer prefers the challenger exactly when $a + b\theta < \bar p$, that is
when
\begin{equation}
\theta \;<\; \theta_{\mathrm{cut}} \;\equiv\; \frac{\bar p - a}{b}.
\label{eq:thetacut}
\end{equation}
Any strictly increasing schedule produces a cutoff of the form
Eq.~\ref{eq:thetacut}, and what fixes its location is the slope's magnitude
rather than any relation to cost: a
\emph{steeper} schedule prices heavy customers out sooner. This is worth
separating from the role of $b < c_1$, which is different and easy to
conflate with it. That condition makes the challenger's margin
$(a - c_0) + (b - c_1)\theta$ decline in $\theta$, so the challenger earns
most on the lightest customers---which is what makes the strategy a skim
rather than an undifferentiated price cut---and it places the challenger's
\emph{least} profitable customer at the cutoff itself.

Profitability across the whole switching set therefore reduces to one check,
at $\theta_{\mathrm{cut}}$. A revenue-neutral schedule satisfies it for
free. Revenue-neutral calibration makes the tenant at reference-mean
consumption pay exactly $\bar p$, so the cutoff sits at the mean tenant and
the challenger's margin on the marginal switcher is
$\bar p - c(\theta_{\mathrm{cut}})$: the incumbent's own margin on its
average customer, independent of $b$, and strictly positive whenever the
incumbent earns a margin there rather than merely covering its costs. The
schedule is self-enforcing---the challenger refuses nobody, and the sorting
costs it nothing.

Absent that property the argument would need
$\theta_{\mathrm{cut}} \le (a - c_0)/(c_1 - b)$, which $b < c_1$ alone does
not deliver: a challenger that undercuts hard on the intercept attracts
customers it cannot serve profitably, and must then ration rather than sort.

\section{Notation}
\label{app:notation}

Table~\ref{tab:notation-full} collects every symbol used in the paper.

\begin{table}[h]
\centering\footnotesize
\begin{tabular}{@{}llp{0.52\columnwidth}@{}}
\toprule
Symbol & Defined & Meaning \\
\midrule
\multicolumn{3}{@{}l}{\emph{Pricing mechanism}}\\
$g$ & \S\ref{sec:price-family} & Hardware-tier index. Calibration is per-tier; this paper instantiates one tier (8$\times$B200). \\
$\bar p_g$ & \S\ref{sec:price-family} & Prevailing flat node-hour rate on tier $g$. Written $\bar p$ in \S\ref{sec:cream}, where one tier is in play. \\
$r$, $D$ & \S\ref{sec:vector} & Resource index and the number of priced dimensions, $D=3$: tensor throughput, HBM bandwidth, NVLink traffic. \\
$u_r(t)$ & \S\ref{sec:vector} & Instantaneous fraction-of-peak utilization of resource $r$; $u(t) \in [0,1]^D$. \\
$\bar u_{g,r}$ & \S\ref{sec:price-family} & \emph{Reference-mix} mean utilization of resource $r$ --- the calibration input. Time-weighted and pooled across the GPUs in a rental (\S\ref{app:stats}). \\
$\bar u^{(i)}_r$ & \S\ref{sec:price-family} & \emph{Tenant $i$'s own} time-averaged utilization. Distinct from $\bar u_{g,r}$, and easy to confuse with it. \\
$\alpha$ & \S\ref{sec:price-family} & Occupancy fraction of the flat rate, and simultaneously the lower clip. $\alpha_g = \alpha\,\bar p_g$. \\
$\beta_{g,r}$ & \S\ref{sec:price-family} & Per-unit price of resource $r$ on tier $g$. \\
$T$ & \S\ref{sec:price-family} & Length of the billing window $[0,T]$. \\
$m(t)$ & --- & Spot multiplier, scaling the whole rate vector uniformly. \\
\midrule
\multicolumn{3}{@{}l}{\emph{Cost and competition}}\\
$\theta$ & \S\ref{app:cream} & A customer's resource intensity. \\
$c(\theta) = c_0 + c_1\theta$ & \S\ref{app:cream} & Cost to serve an hour at intensity $\theta$; $c_1 > 0$ the intensity-dependent slope. \\
$a + b\theta$ & \S\ref{app:cream} & A challenger's consumption-price schedule; $b < c_1$ makes its margin decline in $\theta$. \\
$\theta_{\mathrm{cut}}$ & \S\ref{app:selfselect} & Intensity above which a customer prefers the flat rate, $(\bar p - a)/b$. \\
\midrule
\multicolumn{3}{@{}l}{\emph{Demand and revenue}}\\
$\varepsilon_L$, $\varepsilon_H$ & \S\ref{sec:breakeven} & Names for the two demand channels: light-segment expansion and heavy-segment churn. Used qualitatively --- no demand system is fitted and no elasticity is estimated (\S\ref{sec:market}). \\
$\Delta R/R$ & \S\ref{sec:market} & Fractional revenue change under FBP, against flat pricing on the same pool. \\
$\kappa$ & \S\ref{app:kappa} & Clipping give-back, $\kappa = \tau - \phi \ge 0$: the entire cost of the predictability band on a fixed base. \\
$\tau$, $\phi$ & \S\ref{app:kappa} & Revenue truncated by the cap and added back by the floor; $\phi = 0$ by construction. \\
$w_i$ & \S\ref{app:market} & Tenant $i$'s willingness to pay as a multiple of the flat rate. Distinct from the spot multiplier $m(t)$. \\
\bottomrule
\end{tabular}
\caption{Complete notation.}
\label{tab:notation-full}
\end{table}

\section{Provider Pricing Survey}
\label{app:survey}

Table~\ref{tab:survey} reports the survey summarized in
\S\ref{sec:motivation-survey}.

\begin{table*}[h]
\centering\small
\begin{tabular}{@{}lcccc@{}}
\toprule
Provider & Granularity & Commitment & Differ. & Consump. \\
\midrule
\multicolumn{5}{@{}l}{\emph{Neoclouds / inference specialists}}\\
CoreWeave  & hr/sec   & od/spot/rsv & HW   & per-res \\
Lambda     & hr/min   & od/rsv      & HW   & per-res \\
RunPod     & per-sec  & od/spot/rsv & HW+WL& per-res \\
Together   & tok/hr   & od/rsv      & HW+WL& tok+res \\
Fireworks  & tok/sec  & od/rsv     & WL/QoS& per-tok \\
Modal      & per-sec  & od only     & HW+rel& per-res \\
\midrule
\multicolumn{5}{@{}l}{\emph{Hyperscalers}}\\
AWS   & sec/tok & od/rsv/spot/sp & HW+silicon & res+tok \\
GCP   & sec/tok & od/cud/spot    & HW+TPU     & res+tok \\
Azure & hr/tok  & od/rsv/spot/ptu& HW+PTU     & res+tok \\
OCI   & per-hr  & od/flex/rsv    & HW        & bundle+char \\
\bottomrule
\end{tabular}
\caption{Pricing-structure survey, ten providers, four axes. Granularity:
hr=hourly, sec/min=per-second/minute, tok=per-token. Commitment:
od=on-demand, rsv=reserved, cud=committed-use, sp=savings plan,
ptu=provisioned throughput. Differ.: HW=hardware-only tiers, WL=workload/QoS
tiers, rel=reliability/region. Consump.: per-res=per-resource (GPU+CPU+RAM
\`a la carte), tok=per-token, char=per-character. The pattern
is \emph{not} a clean neocloud/hyperscaler split: per-resource and
per-token billing appear on both sides, and providers cluster by
\emph{product type} (raw-GPU rental vs.\ managed inference) rather than by
camp.}
\label{tab:survey}
\end{table*}

The survey documents pricing \emph{structure}, not revenue: list-price
granularity does not reveal how much revenue flows through each mode. Nor
is the neocloud list-price undercut fully quality-adjusted---SLA,
interconnect and reliability differences, and in part debt-financed
hardware, account for some of the gap. The claim it supports is directional:
the market has moved from time toward consumption, per-resource pricing
already exists in the flexible segment, and that is the direction Agora
extends.

\section{Market-Experiment Construction}
\label{app:market}

This appendix specifies the microsimulation of \S\ref{sec:market}. It fits
no demand system and estimates no elasticity; it prices a pool of measured
configurations under two regimes and counts the difference.

\paragraph{The pool.} Each measured serving configuration---a model at a
given batch size and context length---is one job, carrying the
resource-utilization vector we measured for it, the duration that
configuration ran for, and hence both a flat charge and an FBP bill under
Eq.~\ref{eq:bill}. Configurations differ in duration, since output context
length sets how long a run takes, so revenue is accumulated in dollars
rather than in job counts throughout; a job served for twice as long
contributes twice as much to either regime's revenue.

\paragraph{Willingness to pay, and who is an incumbent.} We express each
job's WTP as a multiple of the flat rate, $\mathrm{WTP}_i = w_i\,\bar p_g$,
so the experiment is scale-free: joining and leaving become statements about
$w_i$ relative to $1$ and to the job's bill ratio, and no absolute dollar
valuation need be assumed. We draw $w_i$ once per job from one of three
families: lognormal in the baseline, a Pareto variant that strengthens the
upper tail, and a bounded uniform on $[c(1-s),\,c(1+s)]$ that removes the
tail altogether. The first two encode the usual prior---most users value
compute near the prevailing rate while a minority value it far above---and
the third deliberately denies it, so the reported result brackets tail
shape rather than assuming it. The uniform is the least favourable of the
three and anchors the low end of the family sweep in \S\ref{sec:market}. No public dataset fixes
this distribution: request-level traces such as the Azure LLM inference
trace~\cite{Azure_2023} record arrivals and token counts but not what a
tenant would have paid, which is why the willingness-to-pay assumption is
swept rather than estimated.

The incumbent set is then not assumed but derived: under flat pricing a job
rents iff $w_i \ge 1$. The $t{=}0$ state is therefore an equilibrium of the
flat regime rather than an exogenous split, which matters because the two
populations then differ only in the way flat pricing itself distinguishes
them. It also removes a free parameter: the relative mass of the two pools
is fixed by the WTP distribution rather than chosen, so the size of the
entry channel is swept implicitly whenever the distribution's parameters
are. The cost of that is a coupling worth stating---a fatter lower tail
means both more prospective demand and a different incumbent mix---so the
sweep reports the two pools' realized masses alongside every point.

\paragraph{Calibration and the switch.} The rate vector is calibrated for
revenue-neutrality over the incumbent set alone (\S\ref{sec:price-family}),
and frozen before any entry or exit occurs. This avoids a fixed point
between prices and population, at the cost of the mechanism being optimal
for a population it will not end up serving---which is the situation a
provider making this switch is actually in. FBP is then applied to the whole
pool: a job rents iff $\mathrm{WTP}_i \ge \mathrm{Bill}_i$. \emph{Leavers}
are incumbents whose premium pushes their bill above their WTP;
\emph{joiners} are non-incumbents whose discount brings their bill below a
WTP the flat rate had excluded. The reported quantity is
$\Delta R/R$: the total FBP revenue of the post-switch served set against
the total flat revenue of the pre-switch served set.

Joiners are absorbed rather than displacing incumbents, which assumes the
provider has slack at the margin. That is the regime in which attracting
light demand is possible at all, and where fleet capacity binds instead,
expansion becomes displacement and this construction does not apply
(\S\ref{sec:market}).

\paragraph{Sweep.} We sweep the WTP distribution rather than fixing it,
along the center and spread that govern entry and churn, and across three
families spanning tail behaviour, reporting $\Delta R/R$ as a region rather
than a point.

Two further axes concern the size of the market rather than the shape of
demand, and are kept separate from the headline for that reason. The
\emph{pool multiplier} scales the price-excluded population beyond what the
WTP distribution implies, modelling latent demand this corpus does not
sample; its instances are drawn conditional on $w_i < 1$, since a job with
$w_i \ge 1$ would already be renting and counting it as a joiner would
double-count demand. The \emph{book share}---the fraction of the addressable
market the provider already serves---is reported as an axis rather than
fixed. Points below a \meas{$5\%$} book share are discarded: $\Delta R/R$ is
normalised by the incumbent book, so a vanishing book makes the ratio
unbounded rather than large, which is an artifact of the normalisation and
not a finding.

Two degenerate cases anchor the sweep. With no entry and no exit the simulation must
return exactly $-\kappa$ (Appendix~\ref{app:kappa}), which validates the
billing pipeline against the calibration and depends on no demand
assumption. And with entry suppressed but churn allowed, $\Delta R/R$ is
strictly below $-\kappa$, which is the pure-downside case a provider should
read first.

\paragraph{Correlation sensitivity.} The one remaining modelling choice we
make explicit is the correlation between WTP and resource intensity:
whether resource-heavy jobs also carry higher valuations, or value compute
independently of how much they consume. Positive correlation dampens
churn---heavy jobs absorb their premium because they value the compute---and
so flatters the mechanism. We take the independent case as the conservative
baseline and report the correlated case as a sensitivity, since it is the
assumption most likely to be contested and the one with the clearest
direction of effect.

\section{Reported Statistics}
\label{app:stats}

The quantities this paper reports are aggregates over a corpus, and several
admit more than one defensible definition. Unless stated otherwise:

\begin{itemize}
\item A workload's \emph{profile} is the time-weighted mean of each $u_r(t)$
over the run, pooled across the GPUs the run occupies. Three choices here
are load-bearing. Time weighting rather than per-sample averaging, because
Eq.~\ref{eq:bill} integrates against time and unequal sample spacing would
otherwise bias the calibration. Pooling across GPUs rather than reporting a
single device, because the bill covers the whole rental and the busiest GPU
in a tensor-parallel job is not representative. And the mean rather than a
median or percentile, because only the mean makes the average tenant pay
exactly $\bar p_g$.

\item \emph{Dispersion} on an axis is the ratio of the largest to the
smallest such profile across the corpus.

\item A workload's \emph{bill ratio} is its Eq.~\ref{eq:bill} charge divided
by the flat charge for the same elapsed time.
\end{itemize}

\section{Rate Calibration}
\label{app:rates}

This appendix derives the rate formula of \S\ref{sec:price-family} and
proves that it delivers the revenue neutrality \S\ref{sec:incentive}
assumes. The construction is an inversion. The intercept already collects
$\alpha\bar p_g$ per unit of rented time, so the usage charges must collect
the remaining $(1-\alpha)\bar p_g$; split equally, each dimension is
responsible for $(1-\alpha)\bar p_g/D$ of it, and since a dimension collects
$\beta_{g,r}\bar u_{g,r}$ from a tenant at the reference mean, setting
$\beta_{g,r}\bar u_{g,r} = (1-\alpha)\bar p_g/D$ and solving gives the
formula. The division by $\bar u_{g,r}$ converts ``dollars this dimension
should collect'' into ``price per unit of utilization.'' What that
construction does not by itself establish is that the rates collect exactly
that from the \emph{mix}, rather than from a hypothetical tenant sitting at
the mean, so we prove it.

\paragraph{Setup.} Index the reference mix by $i$. Tenant $i$ rents for
$T_i$ and consumes
\[
U^{(i)}_r \;=\; \int_0^{T_i} u^{(i)}_r(t)\,dt
\]
of resource $r$ over that rental, so its own time-averaged utilization is
$\bar u^{(i)}_r = U^{(i)}_r/T_i$. Write $T = \sum_i T_i$ for total rented
time and $U_r = \sum_i U^{(i)}_r$ for the mix's total consumption of
resource $r$. The reference mean is the \emph{pooled} average,
\[
\bar u_{g,r} \;=\; U_r / T ,
\]
total consumption over total rented time. This is what
\S\ref{sec:price-family} and Appendix~\ref{app:stats} intend by
``time-weighted, pooled across every GPU in the rental,'' and the proof is
where that definition earns its keep.

\paragraph{Neutrality.} Because Eq.~\ref{eq:bill} integrates an affine rate
over the rental, tenant $i$'s pre-clip bill is
$\alpha_g T_i + \sum_r \beta_{g,r} U^{(i)}_r$. Summing over the mix,
\[
R_{\mathrm{FBP}} \;=\; \alpha\,\bar p_g T \;+\; \sum_r \beta_{g,r} U_r ,
\]
to be compared against $R_{\mathrm{flat}} = \bar p_g T$. Substituting
$\beta_{g,r} = (1-\alpha)\bar p_g/(D\,\bar u_{g,r}) =
(1-\alpha)\bar p_g T/(D\,U_r)$ gives, for every $r$,
\[
\beta_{g,r}\,U_r \;=\; \frac{(1-\alpha)\,\bar p_g\,T}{D} .
\]
The mix's consumption cancels exactly: each dimension collects precisely
$1/D$ of the usage budget from the mix as a whole, whatever that mix
happened to consume. The $D$ usage terms therefore collect
$(1-\alpha)\bar p_g T$ and the intercept collects $\alpha\bar p_g T$, so
$R_{\mathrm{FBP}} = \bar p_g T = R_{\mathrm{flat}}$. The result is exact
rather than approximate, and it is not asymptotic in the size of the mix.

\paragraph{The weighting is load-bearing.} That cancellation needs
$\bar u_{g,r}$ to be $U_r/T$ specifically. Suppose one used the mean of
per-rental means instead, $\tilde u_r = N^{-1}\sum_i \bar u^{(i)}_r$, which
weights a twenty-second job equally with a twenty-minute one. Then
\[
R_{\mathrm{FBP}} \;=\; \bar p_g T\Bigl[\alpha \;+\;
\tfrac{1-\alpha}{D}\textstyle\sum_r \bar u_{g,r}/\tilde u_r\Bigr],
\]
which is neutral only if those $D$ ratios average to one. They do not in
general. Where rental duration correlates positively with utilization on
axis $r$, the pooled mean exceeds the mean of means, $\beta_{g,r}$ comes out
too large, and the provider over-collects on that axis; where the
correlation is negative it under-collects.

This is a live distinction for the corpus of \S\ref{sec:method} rather than
a technicality. Rental duration there is set by output context length, and
long-context runs are decode-dominated---heavy on HBM bandwidth, light on
the tensor engine---so duration correlates positively with one priced axis
and negatively with another at the same time, and the two errors would not
cancel.

\paragraph{Two things neutrality does not depend on.} It does not depend on
equal weighting. For any weights $w_r \ge 0$ with $\sum_r w_r = 1$, setting
$\beta_{g,r} = w_r(1-\alpha)\bar p_g/\bar u_{g,r}$ gives
$\beta_{g,r}U_r = w_r(1-\alpha)\bar p_g T$, which sums to the same
$(1-\alpha)\bar p_g T$. The weights fix how the usage budget is divided
across dimensions, and hence how bills redistribute across tenants, but not
how large that budget is; the choice argued for in \S\ref{sec:price-family}
is revenue-free by construction.

Nor does it depend on $\alpha$. Every $\alpha \in [0,1]$ is neutral, because
the intercept collects $\alpha$ of the total and the usage terms collect
$1-\alpha$ of it by construction. $\alpha$ trades occupancy charge against
usage charge without changing what the provider collects, which is the
precise sense in which raising it blunts the consumption signal at no
revenue gain.

\paragraph{After clipping.} All of the above is pre-clip. Clipping truncates
the largest bills and cannot add revenue back, since the floor never binds
(Appendix~\ref{app:kappa}), so realized revenue on a fixed base is
$R_{\mathrm{flat}} - \kappa$ exactly. That is the whole of the gap between
the neutrality proved here and the neutrality a deployed mechanism
realizes.

\section{Generative AI Use Disclosure}
Claude (Opus 5) was utilized to generate sections of this Work, including text, tables, graphs, and code.
\end{document}